\documentclass[superscriptaddress,twocolumn,amsmath,amssymb]{revtex4}
\usepackage{graphicx}
\usepackage{color}

\renewcommand{\vec}[1]{\mbox{\boldmath $#1$}}

\begin{document}

\title{Spatial correlation of a particle-hole pair 
with a repulsive isovector interaction} 

\author{K. Hagino}
\affiliation{ 
Department of Physics, Kyoto University, Kyoto 606-8502, Japan} 

\author{H. Sagawa}
\affiliation{
RIKEN Nishina Center, Wako 351-0198, Japan}
\affiliation{
Center for Mathematics and Physics,  University of Aizu, 
Aizu-Wakamatsu, Fukushima 965-8560,  Japan}


\begin{abstract}
We study the spatial correlation of a particle-hole pair 
in the isovector channel in $^{56}$Co and $^{40}$K nuclei. 
To this end, we employ 
the Hartree-Fock+Tamm-Dancoff approximation with the Skyrme interaction. 
We find a large concentration of the two-body density 
at positions where 
the neutron particle and the proton hole states locate on the opposite 
side to each other with respect to the core nucleus. 
This feature originates from a repulsive nature of the isovector residual 
interaction, which is in stark contrast to the dineutron correlation 
with an attractive pairing interaction between the valence neutrons 
discussed e.g., in $^{11}$Li and $^6$He.  A possible experimental 
implication of the repulsive correlation is also discussed. 
\end{abstract}


\maketitle

\section{Introduction}

It has been well recognized that the pairing correlation among 
valence neutrons plays a decisive role in 
the structure of weakly bound nuclei \cite{Doba96,BE91,EBH97,HS05,barranco01}. 
In particular, 
there have been several theoretical studies of 
a strong di-neutron correlation \cite{HS05,HTS13,HSCS07,MMS05,PSS07,Zhukov93}, 
in which two neutrons attract each other and show a large probability 
of a two-body wave function with a small 
correlation angle in the coordinate space. 
A strong signature of the dineutron correlation has also been found experimentally 
in several weakly bound nuclei such as $^{11}$Li and $^{6}$He 
\cite{N06,A99,Kubota20}, and more recently in $^{19}$B \cite{Cook20}. 

It would be an interesting question to ask what happens 
to the spatial correlation 
when the interaction is 
repulsive rather than attractive. This could be learned from  
atomic physics, in which the primary interaction among electrons is the Coulomb repulsion. 
It has been actually known that the anti-correlation, opposite to the dineutron correlation, 
exists in the spatial distribution of two electrons in e.g., 
He atoms \cite{MP63, RB79,BY82}. This anti-correlation is referred to as the Coulomb 
hole, in which the correlation angle between two electrons is 
almost 180$^\circ$ in the spatial distribution. 
The purpose of this paper is to address to what extent a similar correlation exists 
in nuclear systems. 

Besides the trivial Coulomb repulsion between protons, several other repulsive 
interactions are known also in nuclear physics.
A well-known example is the isovector particle-hole ($p$-$h$) interaction, 
which plays an important role in generating a collectivity of giant dipole 
resonances (GDR). The repulsive nature of the interaction is evidenced in the fact 
that the empirical mean GDR energy scales as $E\sim 80 A^{-1/3}$ MeV, where $A$ is 
the mass number of a nucleus, while a typical energy for unperturbed 1-particle-1-hole (1p-1h) 
states is given by $E\sim 41 A^{-1/3}$ MeV \cite{RS80,Rowe2010}. 
This suggest that a similar anti-correlation to the Coulomb hole may be seen 
in nuclear systems as well when one considers a particle-hole pair with a 
proton and a neutron. 

In this paper, we pursue this possibility 
by studying nuclei with one neutron particle 
and one proton hole on top of a doubly magic nucleus, 
such as $^{56}$Co (=$^{58}$Ni+n-p) and $^{40}$K (=$^{40}$Ca+n-p). 
To construct the density distribution for a particle and a hole states, 
we first obtain the ground state of the double magic nuclei in the Hartree-Fock 
approximation, and then linearly superpose several 1p-1h states of a neutron-proton pair. 
The coefficients of the superposition is determined by diagonalizing the many-body 
Hamiltonian with the residual interaction. 
This approach is nothing but the Hartree-Fock (HF) plus Tamm-Dancoff 
Approximation (TDA) \cite{RS80,Rowe2010}. 
Of course, one can take into account the ground state correlation of the core 
nuclei within the Random Phase Approximation (RPA). 
However, we prefer the simpler approach 
of TDA, partly because the most of previous three-body model calculations for 
neutron-rich nuclei are based on the particle-particle TDA. 
We believe that a qualitative feature of the spatial distribution does not change 
significantly even when one employs RPA instead of TDA. 

The paper is organized as follows. 
In Sec. II, we first detail the HF+TDA method for neutron-proton particle-hole 
configurations. 
In Sec. III, we apply the HF+TDA method to the ground and excited states of 
$^{56}$Co and $^{40}$K, and discuss the spatial correlation of the isovector 
particle-hole pair. 
We then summarize the paper in Sec. IV. 

\section{Hartree-Fock+Tamm-Dancoff Approximation}

We consider a nucleus with a neutron particle  and a proton hole outside 
a double magic nucleus. To describe such nucleus, we first obtain the ground 
state of the doubly magic nucleus in the Hartree-Fock approximation. 
This procedure also defines the creation operators $a_p^\dagger$ for 
(neutron) particle states and $b_h^\dagger$ for (proton) 
hole states. We superpose several 
particle-hole pairs as 
\begin{equation}
|\Psi\rangle=\sum_{p,h}C_{ph}|ph^{-1}\rangle, 
\label{eq:TDA}
\end{equation}
with 
\begin{equation}
|ph^{-1}\rangle=[a_p^\dagger b_h^\dagger]|0\rangle, 
\end{equation}
where $|0\rangle$ is the ground state of the core nuclei. 
The coefficients $C_{ph}$ are determined by diagonalizing a many-body Hamiltonian, 
whose matrix elements read \cite{RS80} 
\begin{eqnarray} 
\langle ph^{-1}|H|p'h'^{-1}\rangle &=& 
(\epsilon_p-\epsilon_h)\delta_{ph,p'h'} 
+\langle ph^{-1}|\bar{v}_{\rm res}|p'h'^{-1}\rangle, \nonumber \\ \\
&=&(\epsilon_p-\epsilon_h)\delta_{ph,p'h'} 
+\langle ph'|\bar{v}_{\rm res}|hp'\rangle,
\end{eqnarray}
where $\epsilon_p$ and $\epsilon_h$ are single-particle energies for 
particle and hole states, respectively, and $\bar{v}_{\rm res}$ is the anti-symmetrized 
residual interaction. 

Since we consider doubly magic 
nuclei and their vicinity, it is reasonable to assume that the nuclei we discuss 
in this paper are all spherical. 
We then introduce the notation \cite{Rowe2010}
\begin{equation}
a_p^\dagger = a_{jm}^\dagger,~~~
b_p^\dagger = b_{jm}^\dagger=(-1)^{j+m}a_{j-m}, 
\label{eq:hole-op}
\end{equation} 
where $j$ is the total single-particle angular momentum and $m$ is its $z$ 
component. We have suppressed the isospin, the orbital 
angular momentum, and the radial quantum numbers 
to simplify the notation, but they should be understood as implicitly specified. 
A particle-hole state with the coupled angular momentum $J$ and its $z$-component $M$ 
then reads, 
\begin{eqnarray}
|ph^{-1}; JM\rangle&=&
\sum_{m_p,m_h}\langle j_p m_p j_h m_h|JM\rangle 
a_{j_pm_p}^\dagger b_{j_hm_h}^\dagger |0\rangle, \nonumber \\ \\
&=&\sum_{m_p,m_h}(-1)^{j_h+m_h} 
\langle j_p m_p j_h m_h|JM\rangle \nonumber \\
&&\times 
a_{j_pm_p}^\dagger a_{j_h-m_h} |0\rangle. 
\end{eqnarray}

The particle-hole wave function in the coordinate space can be 
constructed as \cite{BBR67} 
\begin{equation}
\Psi(\vec{r}_p,\vec{r}_h)=\sum_{p,h}C_{ph}
\Psi_{ph}(\vec{r}_p,\vec{r}_h), 
\end{equation}
with 
\begin{eqnarray}
\Psi_{ph}(\vec{r}_p,\vec{r}_h) 
&=&
\sum_{m_p,m_h}(-1)^{j_h+m_h} 
\langle j_p m_p j_h m_h|JM\rangle \nonumber \\
&&\times 
\langle \vec{r}_p|\phi_{j_pl_pm_p}\rangle 
\langle \phi_{j_hl_h-m_h}|\vec{r}_h\rangle, 
\end{eqnarray}
where $\langle\vec{r}|\phi_{jlm}\rangle=\phi_{jlm}(\vec{r})$ 
is a single-particle wave function, with $l$ being the orbital angular momentum. 
The density distribution is then obtained as 
\begin{equation}
\rho(\vec{r}_p,\vec{r}_h) 
=\sum_{m_s,m_{s'}}|\langle \chi_{m_s}|\Psi(\vec{r}_p,\vec{r}_h)|\chi_{m_{s'}}\rangle|^2, 
\end{equation}
where $|\chi_m\rangle$ is the spin wave function. 

\section{Spatial correlation of an neutron-proton particle-hole pair}

Let us now apply the method presented in the previous section to actual nuclei. 
We first discuss the $^{56}$Co nucleus, which can be viewed as the 
doubly magic nucleus $^{56}$Ni with one neutron particle and one proton hole. 
For this purpose, we use the Skyrme functional \cite{VB72}. 
While we employ the full Skyrme 
functional for the ground state of $^{56}$Ni, we use only the $t_0$ and $t_3$ 
terms of the residual interaction for the TDA calculations. 
That is, we use the isovector residual interaction given by \cite{HS99} 
\begin{eqnarray}
\bar{v}_{\rm res}(\vec{r},\vec{r}') 
&=&-\left[\frac{t_0}{4}(1+2x_0)+\frac{t_3}{24}(1+2x_3)\rho(\vec{r})^\alpha
\right]\vec{\tau}\cdot\vec{\tau}' \nonumber \\
&&\times \delta(\vec{r}-\vec{r}'), 
\end{eqnarray}
where $\rho(\vec{r})$ is the total density and $\vec{\tau}$ is the isospin 
operator. 
$t_0$, $t_3$, $x_0$, $x_3$, and $\alpha$ are parameters in the Skyrme interaction. 
Notice that the matrix element of the 
$\vec{\tau}\cdot\vec{\tau}'$ factor is 
$\langle pn|\vec{\tau}\cdot\vec{\tau}'|np\rangle=2$ for a neutron-proton particle-hole 
pair. 
For simplicity, we neglect the term which is proportional to 
$(\vec{\sigma}\cdot\vec{\sigma}')(\vec{\tau}\cdot\vec{\tau}')$ in the residual 
interaction, and thus we focus only on natural parity states. 
The matrix elements of the residual interaction may be easily evaluated by 
using the helicity representation as in Ref. \cite{BE91}. 

\begin{figure}[tb]
\begin{center}
\includegraphics[
clip,width=0.9\columnwidth]{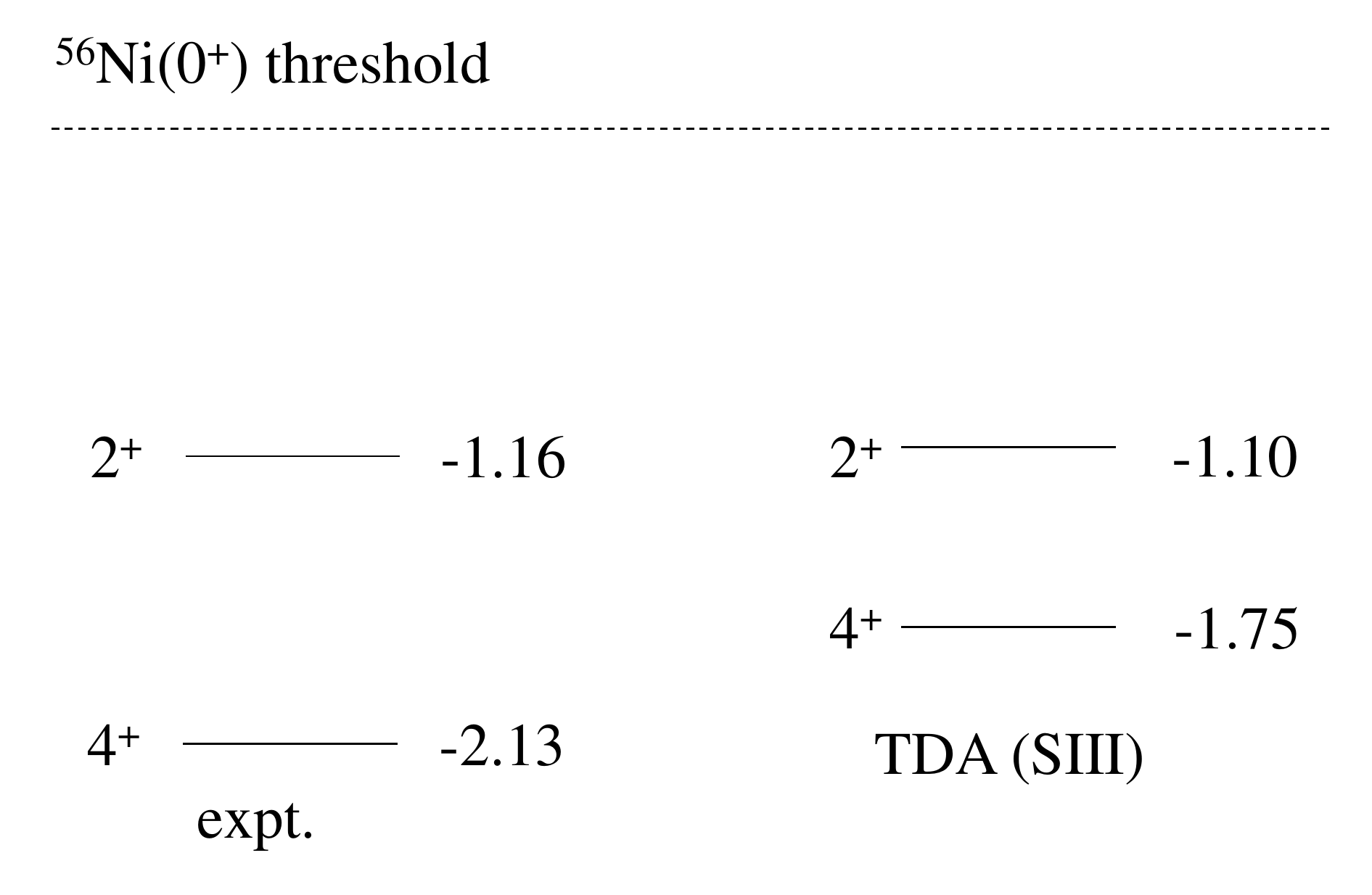}
\caption{
The energies of the first 4$^+$ and 2$^+$ states in $^{56}$Co calculated with the 
neutron-particle proton-hole Tamm-Dancoff Approximation (TDA) 
with the Skyrme SIII interaction. 
These energies are measured from the ground state of $^{56}$Ni after 
correcting the mass difference between a neutron and a proton. 
} 
\end{center}
\end{figure}

In this paper, we mainly employ the SIII interaction \cite{SIII}. 
We have confirmed that the results are not qualitatively altered even if we 
use other parameter sets of the interaction, such as SLy4 \cite{SLy4}. 
The continuum spectrum for neutron particle states is discretized with the box boundary 
condition with the box size of 15 fm, and is truncated at $\epsilon_p=40$ MeV. 
The calculated energies of the first natural-parity 
4$^+$ and 2$^+$ states in $^{56}$Co 
are shown in Fig. 1. These energies are given with respect to the ground 
state of $^{56}$Ni, after correcting the mass difference between a proton and a neutron. 
Even though our aim is not to reproduce the experimental spectrum, the agreement 
with the experimental data is satisfactory. 
The components of the wave functions for these states, that is, 
$|C_{ph}|^2$ in Eq. (\ref{eq:TDA}), are summarized in Table I. 
Since  
the single-particle levels up to 
$1f_{7/2}$ are fully occupied, and the lowest unoccupied level is 
$2p_{3/2}$, in the core nucleus $^{56}$Ni, 
the wave functions are dominated by 
the $(2p_{3/2})_n (1f_{7/2})_p^{-1}$ configuration, even though there 
are appreciable mixtures of other components as well. 

\begin{table}[tb]
\begin{center}
\begin{tabular}{c|cc}
\hline
\hline
components & $4_1^+$ & $2_1^+$ \\
\hline
$(2p_{3/2})_n (1f_{7/2})_p^{-1}$  & 96.4 & 95.5   \\
$(2p_{1/2})_n (1f_{7/2})_p^{-1}$  & 2.85 &  --  \\
$(1f_{5/2})_n (1f_{7/2})_p^{-1}$  & 0.531 & 3.26   \\
$(1h_{11/2})_n (1f_{7/2})_p^{-1}$  & 0.00557  & 0.309   \\
$(1g_{9/2})_n (1d_{5/2})_p^{-1}$  &  1.83$\times 10^{-5}$  & 0.202   \\
\hline
\hline
\end{tabular}
\caption{
The components of the wave functions in percent 
for the first 4$^+$ and 2$^+$ states 
in $^{56}$Co obtained with the neutron-particle proton-hole Tamm-Dancoff 
Approximation. 
}
\end{center}
\end{table}

\begin{figure}[tb]
\begin{center}
\includegraphics[
clip,width=0.6\columnwidth]{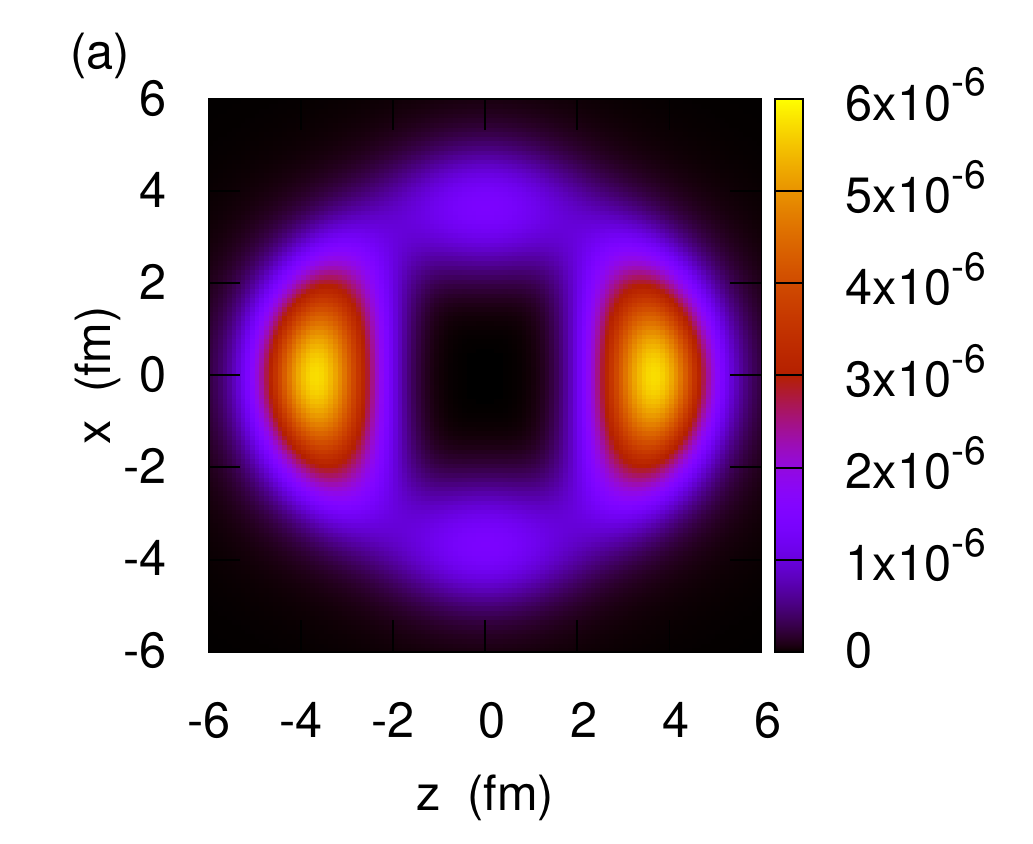}
\includegraphics[
clip,width=0.6\columnwidth]{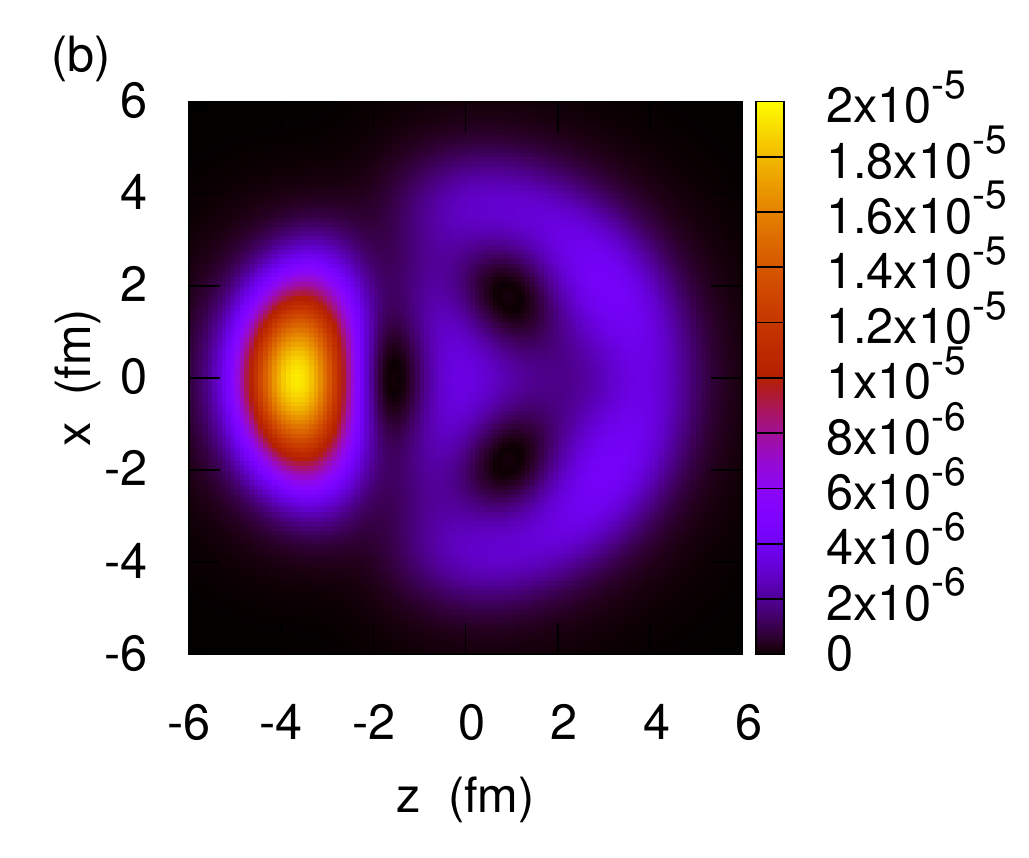}
\caption{The uncorrelated (the upper panel) and the correlated (the lower panel) 
proton-hole distributions in the 2-dimensional $(z,x)$ plane 
for the 4$^+$ state in $^{56}$Co, when 
the neutron-particle is located at $(z,x)=(3.7,0.0)$ fm. 
The azimuthal angular momentum component is set to be $M$=0. 
} 
\end{center}
\end{figure}

The spatial distribution for the proton hole in the 2-dimensional 
$(z,x)$ plane is shown in Fig. 2 
for the 4$^+$ state in $^{56}$Co 
with the azimuthal angular momentum component $M$=0. 
To draw the figures, the location of the reference neutron particle state 
is fixed at $(z,x)=(3.7,0.0)$ fm. 
The upper panel shows the unperturbed case with the 
$(1f_{7/2})_p^{-1}$ wave function. 
As expected, the hole wave function $(1f_{7/2})_p^{-1}$ has two symmetric peaks 
at the positions opposite to the center of the core nucleus. 
A similar feature has been known in an uncorrelated two-neutron density 
distribution \cite{HSS2010}. 
The position of the reference neutron-particle state is in fact 
chosen at a place where 
the unperturbed hole distribution takes the maximum. 
The correlated hole density, in which the particle-hole repulsive interaction 
is active, is shown in the lower panel of Fig. 2. 
One can see a strong repulsive correlation, with which the component 
close to the reference neutron-particle state is largely hindered. 
This is in analogous to the Coulomb hole observed in many-electron systems, 
and is completely opposite to the di-neutron configuration, in which the 
two valence neutrons stay mainly 
at the same side of the two-dimensional plane with a small relative distance, 
that is, a small correlation angle. 

\begin{figure}[tb]
\begin{center}
\includegraphics[
clip,width=0.6\columnwidth]{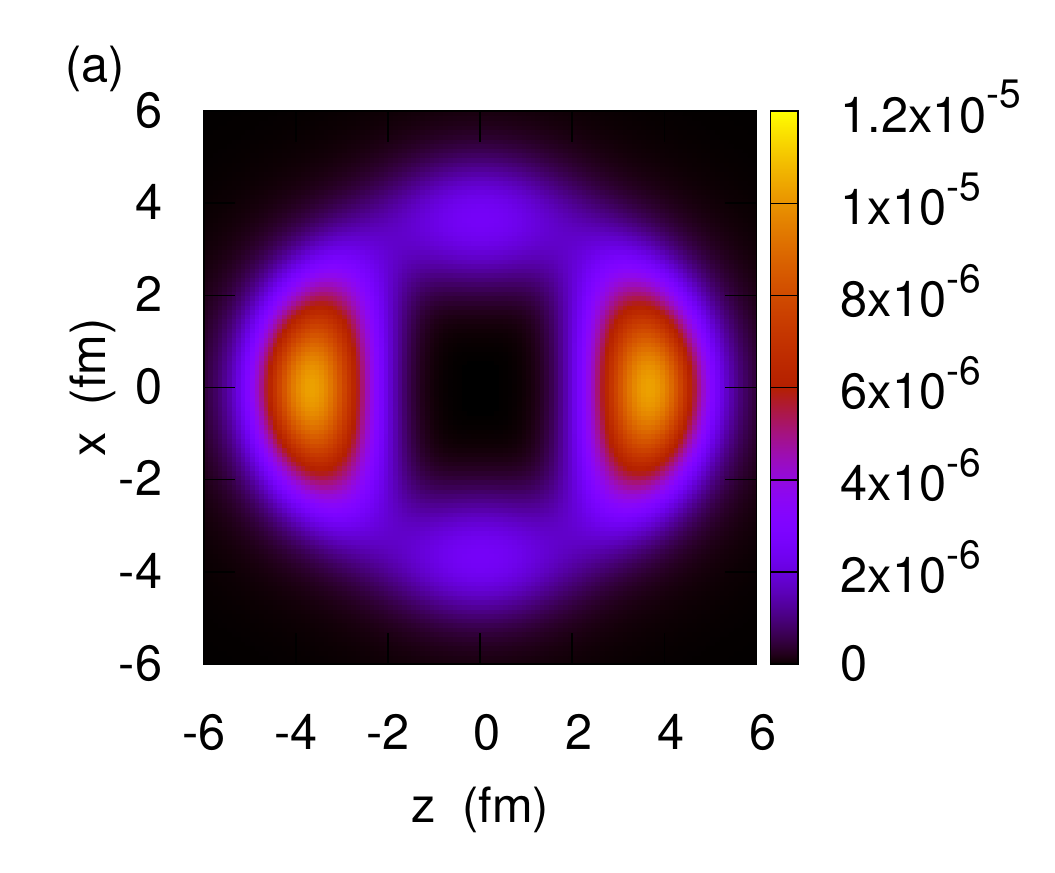}
\includegraphics[
clip,width=0.6\columnwidth]{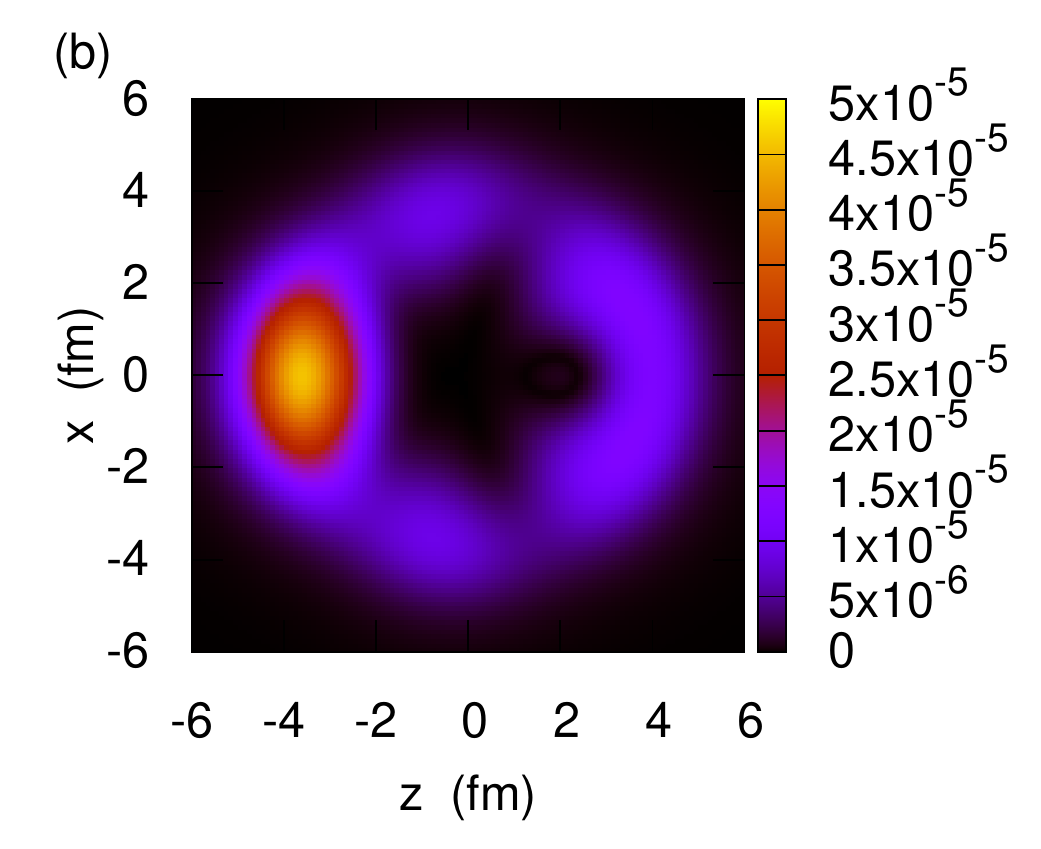}
\caption{The same as Fig. 2, but for the $2^+$ state in 
$^{56}$Co. 
} 
\end{center}
\end{figure}

Fig. 3 shows the spatial hole distribution for the 2$^+$ state in $^{56}$Co. 
General features are quite similar to those for the 4$^+$ state; 
a strong concentration of the hole distribution at the opposite side 
of the reference neutron-particle 
with a small component of the hole distribution at the near side. 
A minor difference is seen in different patterns of the distributions 
at the center of the nucleus: almost no distribution for the 2$^+$ state, 
while an appreciable component exists in the case of the 4$^+$ state.  
Besides this, the strong repulsive correlation can be seen both in the 2$^+$ and 
the 4$^+$ states. 

\begin{figure}[tb]
\begin{center}
\includegraphics[
clip,width=0.6\columnwidth]{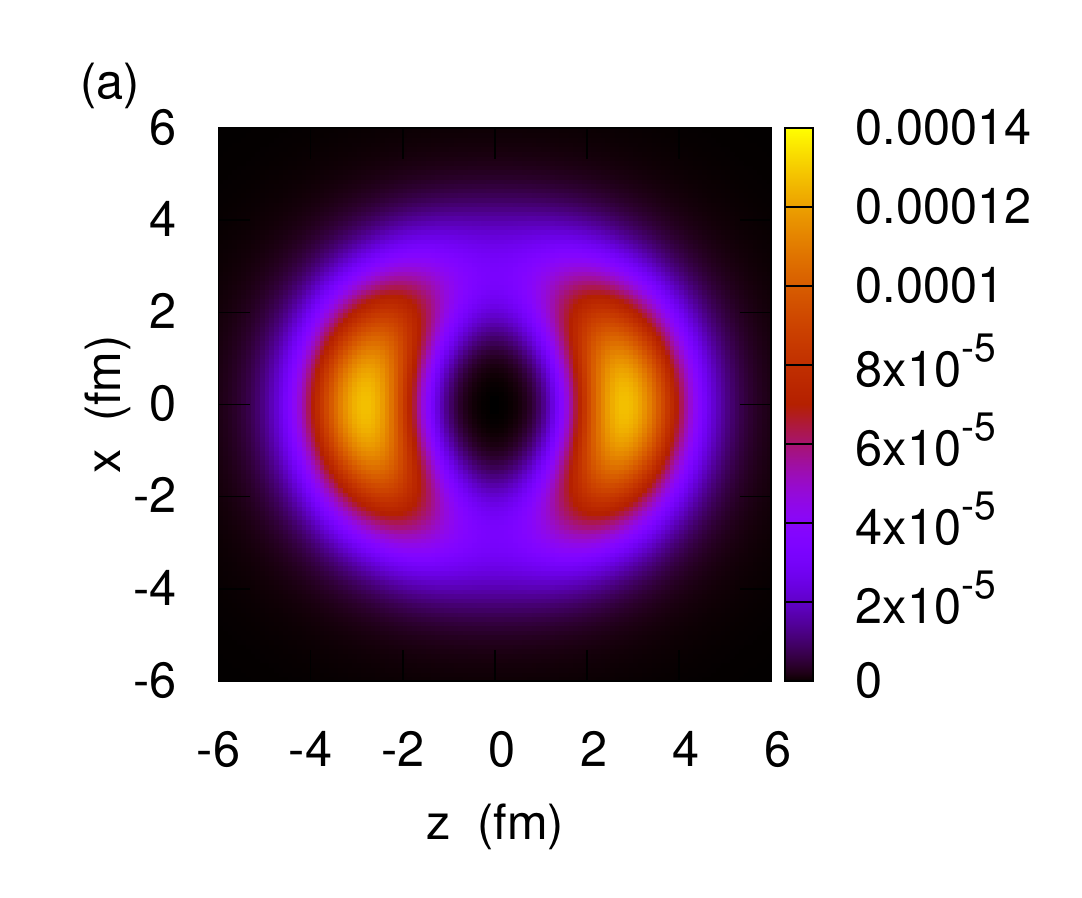}
\includegraphics[
clip,width=0.6\columnwidth]{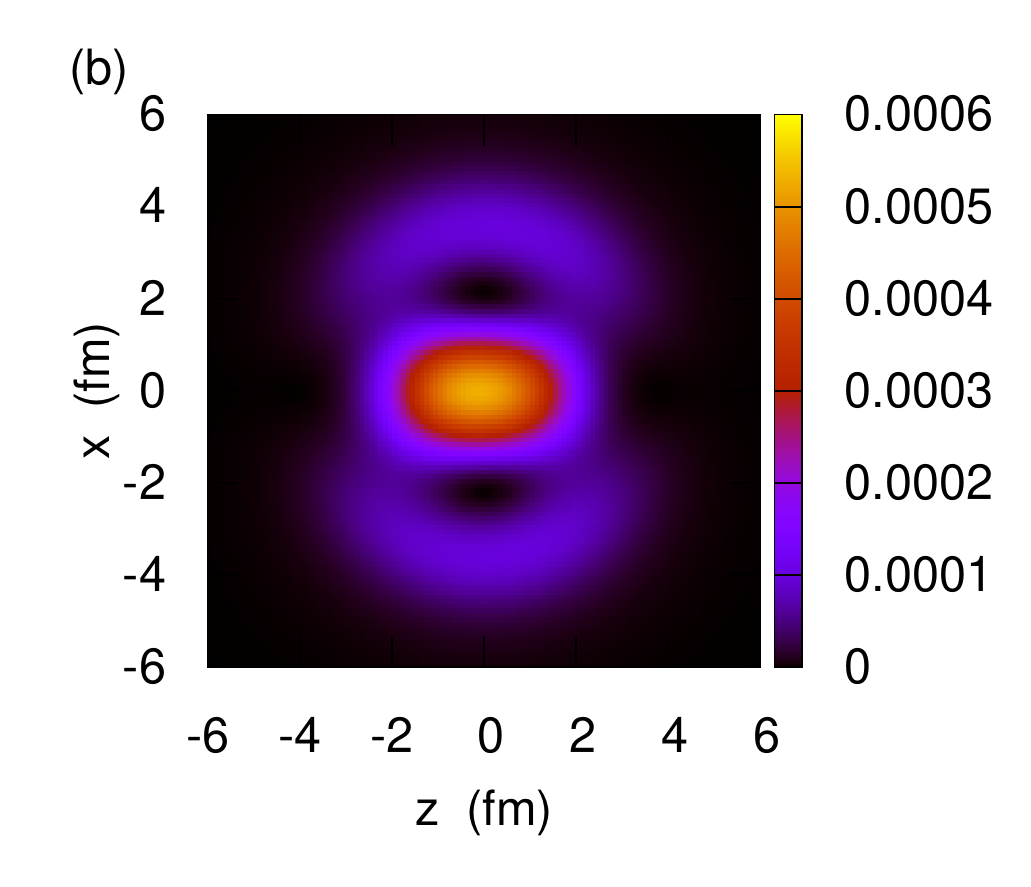}
\caption{The same as Fig. 2, but for the $3^-$ state in 
$^{40}$K when the reference neutron-particle is fixed at 
$(z,x)=(2.8,0.0)$ fm. 
} 
\end{center}
\end{figure}

Let us next discuss the first 3$^-$ state in $^{40}$K, assuming the $^{40}$Ca as a 
core nucleus. The ground state of $^{40}$K is $4^-$, and the $3^-$ is the lowest 
natural parity state. The TDA with the SIII interaction yields the energy of 
the 3$^-$ state to be +0.052 MeV above the ground state of $^{40}$Ca after correcting 
the neutron-proton mass difference, whereas the empirical energy of the 3$^-$ state is 
$-1.28$ MeV. Even though the 3$^-$ state appears above the threshold in the present calculation, 
the TDA energy itself is $-1.24$ MeV for this state before introducing the neutron-proton 
mass difference, and thus the treatment of continuum state 
would not play a crucial role. The wave function of the 3$^-$ state consists of 
81.8\% of the $(1f_{7/2})_n (1d_{3/2})_p^{-1}$ configuration, 
17.9\% of the $(1f_{7/2})_n (2s_{1/2})_p^{-1}$ configuration, 
and 0.112\% of the $(1f_{7/2})_n (1d_{5/2})_p^{-1}$ configuration. 
The density distribution of the proton hole state is shown in Fig. 4, for 
the neutron-particle state fixed at $(z,x)=(2.8,0.0)$ fm. 
Even though the correlation is less pronounced as compared to the $^{56}$Co 
nucleus due to the s-wave component which has a finite value at the origin, 
one can see that the density in the vicinity of the reference neutron-particle is 
largely suppressed, reflecting the repulsive correlation between the neutron-particle 
and the proton-hole. 

\section{Summary}

We have discussed the spatial correlation of an isovector particle-hole pair 
in $^{58}$Co and $^{40}$K 
by using the Hartree-Fock plus Tamm-Dancoff approximation with a Skyrme interaction. 
We have found 
a large concentration of the hole state distribution on the opposite side of 
the reference neutron-particle 
due to the repulsive nature of the isovector residual particle-hole interaction, 
as in the phenomenon of a Coulomb hole in atomic physics. 
This is 
in stark contrast to the dineutron correlation in neutron-rich nuclei, which 
is originated from an attractive 
pairing interaction between valence neutrons. 
This feature has been qualitatively argued in Ref. \cite{B67}, but it had not yet been 
demonstrated in actual numerical calculations. 

In Ref. \cite{B67}, it was argued that the repulsive correlation of an isovector 
particle-hole pair leads to a suppression of a ground-state-to-ground-state 
deuteron transfer reaction, e.g., $^{54}$Fe ($^3$He,p) $^{56}$Co. 
Notice that the two proton holes in $^{54}$Fe prefer the spatial configuration 
in which two holes are close to each other (see also the Appendix below). 
If one of those proton holes is filled in via a deuteron transfer, 
the neutron in the deuteron and the other proton hole would be located close 
to each other. This would correspond to 
an excited state of $^{56}$Co, and thus, the transfer to the ground state of 
$^{56}$Co would be largely suppressed. 
The experimental data shown in Ref. \cite{B67} seem to be consistent with this 
picture. Nevertheless, this has yet to be confirmed with an appropriate reaction 
theory for deuteron transfer reactions. It would be an interesting future work 
to estimate transfer cross sections with the coupled-reaction-channel method 
or the second order distorted wave Born approximation using the 
particle-hole wave functions obtained in this paper. 

\section*{Acknowledgments}
We thank K. Yoshida for useful discussions. 
This work was completed during the long term workshop 
``Mean-field and Cluster Dynamics in Nuclear Systems 2022 (MCD2022)'' 
held at Yukawa Institute for Theoretical Physics, Kyoto University. 
We thank Yukawa Insitute for its hospitality. 
This work was partly supported JSPS KAKENHI Grant Numbers 19K03861 and JP21H00120. 

\appendix
\section{Hole-hole correlation}

Using the operator $b_{jm}^\dagger$ 
defined by Eq. (\ref{eq:hole-op}) for hole states, the structure of a nucleus 
with two hole states from a core nucleus can be described in a similar way as 
a three-body model for two-particle states \cite{BE91,EBH97,HS05}. 
The only difference is that single-particle energies for hole states have to 
be multiplied by a factor of $-1$, as they represent removal energies of a 
particle from a core nucleus. 

We apply this formalism to the $^{54}$Fe nucleus, which can be viewed as a two 
proton hole state from the $^{56}$Ni nucleus. To this end, we use a simple 
zero-range pairing interaction between the proton holes, in which the 
effect of Coulomb repulsion is effectively 
mocked up by adjusting the strength \cite{OHS10}. 
We use the Skyrme Hartree-Fock method with the SIII interaction to generate 
proton single-particle states in $^{56}$Ni. By including all the proton hole 
states, the strength of the pairing interaction is adjusted to be 
$g=262$ MeV fm$^3$ to reproduce the empirical two-proton separation energy 
of $^{56}$Ni, $S_{2p}=12.23$ MeV. 

\begin{figure}[tb]
\begin{center}
\includegraphics[
clip,width=0.7\columnwidth]{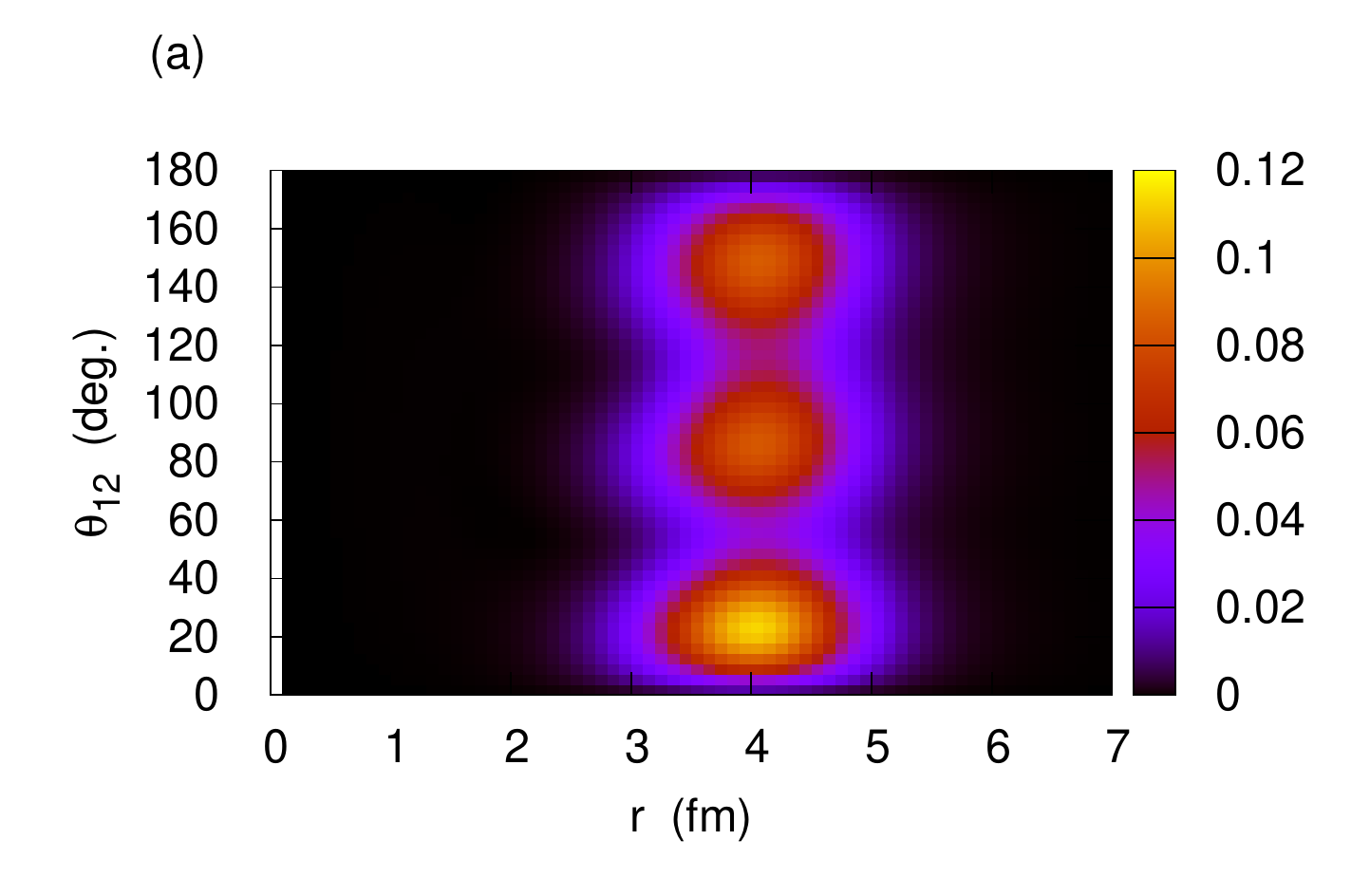}
\includegraphics[
clip,width=0.6\columnwidth]{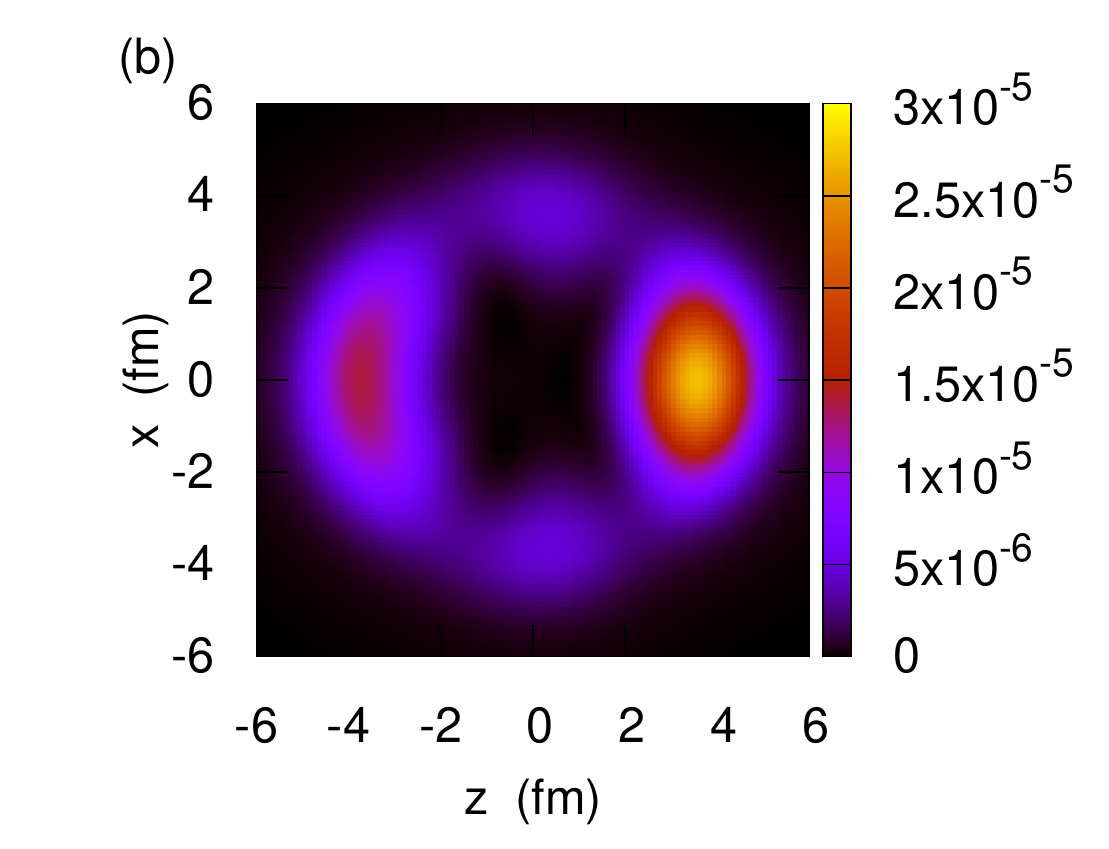}
\caption{
The distribution of the two-hole configuration in $^{56}$Ni obtained 
with the hole-hole Tamm-Dancoff approximation. 
The upper panel shows the density distribution for the two hole states 
obtained by setting $r_1=r_2=r$. This is plotted as a function of $r$ and 
the opening angle, $\theta_{12}$. 
The weight factor $8\pi^2r^4\sin\theta_{12}$ has been multiplied. 
The lower panel shows the density of the second hole when the first hole is 
fixed at $(z,x)=(3.7,0.0)$ fm. 
} 
\end{center}
\end{figure}

The resultant two-hole wave function consists of 
98.5\% of the $[(1f_{7/2}^{-1})^2]$ configuration, 
0.748\% of the $[(1d_{3/2}^{-1})^2]$ configuration, 
0.433\% of the $[(1d_{5/2}^{-1})^2]$ configuration, 
and 0.219\% of the $[(2s_{1/2}^{-1})^2]$ configuration. 
The density distribution of the two-hole state is shown in Fig. 5. 
The upper panel corresponds to the density in the 2-dimensional 
$(r,\theta_{12})$ plane, in which $r_1=r_2=r$ is the distance of the hole 
states from the center of the nucleus and $\theta_{12}$ is the opening 
angle between the two holes. The weight factor of $8\pi^2r^4\sin\theta_{12}$ 
has been taken into account \cite{HS05}.  
Notice a large two-hole probability at small opening angles around 
$\theta\sim 20^\circ$. 
The lower panel, on the other hand, shows the distribution of the second hole 
when the first hole is fixed at $(z,x)=(3.7,0.0)$ fm. 
One can clearly see an enhancement of the density distribution when the two 
hole states are located at similar places. This is basically the same phenomenon 
as the dineutron 
correlation discussed in Ref. \cite{HS05}.

\end{document}